\magnification \magstep1
\raggedbottom
\openup 4\jot
\voffset6truemm
\headline={\ifnum\pageno=1\hfill\else
\hfill {\it Rarita-Schwinger Potentials in Quantum Cosmology}
\hfill \fi}
\def\cstok#1{\leavevmode\thinspace\hbox{\vrule\vtop{\vbox{\hrule\kern1pt
\hbox{\vphantom{\tt/}\thinspace{\tt#1}\thinspace}}
\kern1pt\hrule}\vrule}\thinspace}
\centerline {\bf RARITA-SCHWINGER POTENTIALS IN QUANTUM COSMOLOGY}
\vskip 1cm
\centerline {GIAMPIERO ESPOSITO$^{1,2}$, GABRIELE GIONTI$^{3}$,
ALEXANDER Yu. KAMENSHCHIK$^{4}$,}
\centerline {IGOR V. MISHAKOV$^{4}$ and GIUSEPPE 
POLLIFRONE$^{5}$}
\vskip 1cm
\centerline {\it ${ }^{1}$Istituto Nazionale di Fisica Nucleare,
Sezione di Napoli}
\centerline {\it Mostra d'Oltremare Padiglione 20, 
80125 Napoli, Italy,}
\centerline {\it ${ }^{2}$Dipartimento di Scienze Fisiche}
\centerline {\it Mostra d'Oltremare Padiglione 19,
80125 Napoli, Italy,}
\centerline {\it ${ }^{3}$Scuola Internazionale Superiore
di Studi Avanzati}
\centerline {\it Via Beirut 2-4, 34013 Trieste, Italy,}
\centerline {\it ${ }^{4}$Nuclear Safety Institute}
\centerline {\it Russian Academy of Sciences}
\centerline {\it 52 Bolshaya Tulskaya, Moscow 113191, Russia,}
\centerline {\it ${ }^{5}$Dipartimento di Fisica, Universit\`a
di Roma ``La Sapienza"}
\centerline {\it and INFN, Sezione di Roma}
\centerline {\it Piazzale Aldo Moro 2, 00185 Roma, Italy}
\vskip 2cm
\leftline {PACS numbers: 04.20.Cv, 04.60.Ds, 98.80.Hw}
\vskip 100cm
\noindent
{\bf Abstract.} This paper studies the two-spinor form of the
Rarita-Schwinger potentials subject to local boundary conditions
compatible with local supersymmetry.
The massless Rarita-Schwinger field equations are studied in 
four-real-dimensional Riemannian backgrounds with boundary. Gauge
transformations on the potentials are shown to be compatible
with the field equations providing the background is Ricci-flat,
in agreement with previous results in the literature.
However, the preservation of boundary conditions under such
gauge transformations leads to a restriction of the gauge freedom.
The recent construction
by Penrose of secondary potentials which supplement the 
Rarita-Schwinger potentials is then applied.
The equations for the secondary potentials, jointly with
the boundary conditions, 
imply that the background four-geometry is further
restricted to be totally flat. 
\vskip 100cm
\leftline {\bf 1. Introduction}
\vskip 1cm
\noindent
Over the last few years, many efforts have been produced to
study locally supersymmetric boundary conditions in perturbative
quantum cosmology.$^{1-7}$ The aim of 
this paper is to perform a complete 
analysis of the corresponding {\it classical} elliptic boundary-value
problems. Indeed, in Ref. 8 it was shown that one possible set of local
boundary conditions for a massless field of spin $s$, 
involving field strengths $\phi_{A...L}$ and ${\widetilde \phi}_{A'...L'}$ 
and the Euclidean normal ${_{e}n^{AA'}}$ to the boundary:
$$
2^{s} \; {_{e}n^{AA'}} ... {_{e}n^{LL'}} \; \phi_{A...L}
= \pm \; {\widetilde \phi}^{A'...L'} 
\; \; \; \; {\rm at} \; \; \; \; {\partial M} \; ,
\eqno (1.1)
$$
can only be imposed in flat Euclidean backgrounds with boundary.
However, such boundary conditions (motivated
by supergravity theories in anti-de Sitter 
space-time,$^{9-10}$ where (1.1) is
essential to obtain a well-defined quantum theory
after taking the covering space of anti-de Sitter)
do not make it possible to relate bosonic
and fermionic fields through the action of 
complementary projection operators at the boundary.$^{2}$ 
For this purpose, one has
to impose another set of local and supersymmetric boundary
conditions, first proposed in Ref. 1. 
These are in general {\it mixed}, and 
involve in particular Dirichlet conditions for the transverse modes of
the vector potential of electromagnetism, a mixture of Dirichlet and
Neumann conditions for scalar fields, and local boundary conditions for
the spin-${1\over 2}$ field and the spin-${3\over 2}$ potential. Using
two-component spinor notation for supergravity,$^{5,11-12}$ the
spin-${3\over 2}$ boundary conditions 
relevant for quantum cosmology take the form$^{4}$
$$
\sqrt{2} \; {_{e}n_{A}^{\; \; A'}} \;
\psi_{\; \; i}^{A}= \pm  
{\widetilde \psi}_{\; \; i}^{A'} 
\; \; \; \; {\rm at} \; \; \; \; \partial M \; .
\eqno (1.2)
$$
With our notation, 
$\Bigr(\psi_{\; \; i}^{A},{\widetilde \psi}_{\; \; i}^{A'}\Bigr)$
are the {\it independent} (i.e. not related by any conjugation) 
spatial components (hence $i=1,2,3$) of the spinor-valued 
one-forms appearing in the action 
functional of Euclidean supergravity.$^{5,11}$

In the light of the results outlined so far, a naturally occurring
question is whether an analysis motivated by the one in 
Ref. 8 can be used to derive 
restrictions on the classical boundary-value problem corresponding
to (1.2). Such a question is of crucial importance for at least
two reasons:
\vskip 0.3cm
\noindent
(i) In the absence of boundaries, extended supergravity theories
are naturally formulated on curved backgrounds with a cosmological
constant.$^{5,9,10}$ Thus, if a local theory in terms of 
spin-${3\over 2}$ potentials and in the presence of boundaries can
only be studied in flat Euclidean four-space, this result would 
make it impossible to consider 
the most interesting supergravity models when a four-manifold 
with boundaries occurs.
\vskip 0.3cm
\noindent
(ii) One of the main problems of the twistor programme for general
relativity lies in the impossibility to achieve a twistorial
reconstruction of (complex) vacuum space-times which are not
right-flat (i.e. such that the Ricci spinor $R_{AA'BB'}$ and 
the self-dual Weyl spinor ${\widetilde \psi}_{A'B'C'D'}$ vanish).
To overcome this difficulty, Penrose has proposed a new definition
of twistors as charges for massless spin-${3\over 2}$ fields in
Ricci-flat Riemannian manifolds (see references in the following
sections). However, since gravitino potentials have been studied also
in backgrounds which are not 
Ricci-flat,$^{9-10}$ one is led to
ask whether the recent Penrose formalism can be applied to
study a larger class of Riemannian four-manifolds with boundary.

For this purpose, we introduce in Sec. 2 the Rarita-Schwinger
potentials with their gauge transformations in Riemannian
background four-geometries. Section 3
derives compatibility conditions from the gauge transformations of
Sec. 2, and from the boundary conditions (1.2). 
Section 4 is devoted to the secondary potentials which
supplement the Rarita-Schwinger potentials in 
Ricci-flat backgrounds. Section 5 studies other
sets of gauge transformations.
Concluding remarks and open problems are presented in Sec. 6.
Relevant details about the two-spinor form of Rarita-Schwinger equations
are given in the appendix.
\vskip 1cm
\leftline {\bf 2. Rarita-Schwinger Potentials 
and their Gauge Transformations}
\vskip 1cm
\noindent
For the reasons described in the introduction, we are here interested
in the independent spatial components $\Bigr(\psi_{\; \; i}^{A},
{\widetilde \psi}_{\; \; i}^{A'}\Bigr)$ of the gravitino field
in Riemannian backgrounds. In terms of the spatial components
$e_{AB'i}$ of the tetrad, and of spinor fields, they can be
expressed as$^{11,13-14}$
$$
\psi_{A \; i}= \Gamma_{\; \; AB}^{C'}
\; e_{\; \; C'i}^{B} \; ,
\eqno (2.1)
$$
$$
{\widetilde \psi}_{A' \; i}= 
\gamma_{\; \; A'B'}^{C} \; 
e_{C \; \; \; i}^{\; \; B'} \; .
\eqno (2.2)
$$
A first important difference with respect to the Dirac form of
the potentials studied in Ref. 8 is that the spinor fields 
$\Gamma_{\; \; AB}^{C'}$ and 
$\gamma_{\; \; A'B'}^{C}$ are no longer symmetric in
the second and third index.$^{14}$ From now on, they will be 
referred to as spin-${3\over 2}$ potentials.
They obey the differential equations 
(see appendix and cf. Refs. 13 and 14)
$$
\epsilon^{B'C'} \; \nabla_{A(A'} \; \gamma_{\; \; B')C'}^{A}
=-3 \Lambda  \; {\widetilde \alpha}_{A'} \; ,
\eqno (2.3)
$$
$$
\nabla^{B'(B} \; \gamma_{\; \; \; B'C'}^{A)}
=\Phi_{\; \; \; \; \; \; \; \; C'}^{ABL'}
\; {\widetilde \alpha}_{L'} \; ,
\eqno (2.4)
$$
$$
\epsilon^{BC} \; \nabla_{A'(A} \; \Gamma_{\; \; B)C}^{A'}
=-3\Lambda \; \alpha_{A} \; ,
\eqno (2.5)
$$
$$
\nabla^{B(B'} \; \Gamma_{\; \; \; BC}^{A')}
={\widetilde \Phi}_{\; \; \; \; \; \; \; \; \; C}^{A'B'L}
\; \alpha_{L} \; ,
\eqno (2.6)
$$
where $\nabla_{AB'}$ is the spinor covariant derivative corresponding
to the curved connection $\nabla$ of the background,
the spinors $\Phi_{\; \; \; \; \; C'D'}^{AB}$ and
${\widetilde \Phi}_{\; \; \; \; \; \; \; CD}^{A'B'}$ 
correspond to the trace-free part of the Ricci tensor, the
scalar $\Lambda$ corresponds to the
scalar curvature $R=24\Lambda$ of the background,
and $\alpha_{A},{\widetilde \alpha}_{A'}$ are a pair of
independent spinor fields, corresponding to the Majorana
field in the Lorentzian regime.
Moreover, the potentials are subject to the gauge transformations
(cf. Sec. 5)
$$
{\widehat \gamma}_{\; \; B'C'}^{A}
\equiv \gamma_{\; \; B'C'}^{A}
+\nabla_{\; \; B'}^{A} \; \lambda_{C'} \; ,
\eqno (2.7)
$$
$$
{\widehat \Gamma}_{\; \; BC}^{A'}
\equiv \Gamma_{\; \; BC}^{A'}
+\nabla_{\; \; B}^{A'} \; \nu_{C} \; .
\eqno (2.8)
$$
A second important difference with respect 
to the Dirac potentials$^{8}$ 
is that the spinor fields $\nu_{B}$ and $\lambda_{B'}$ are no
longer taken to be solutions of the Weyl equation.
They should be freely specifiable (see Sec. 3).
\vskip 1cm
\leftline {\bf 3. Compatibility Conditions}
\vskip 1cm
\noindent
Our task is now to derive compatibility conditions, by requiring
that the field equations (2.3)-(2.6) should also be satisfied by the
gauge-transformed potentials appearing on the left-hand side of
Eqs. (2.7)-(2.8). For this purpose, after defining the
operators
$$
\cstok{\ }_{AB} \equiv \nabla_{M'(A} 
\; \nabla_{B)}^{\; \; \; M'} \; ,
\eqno (3.1)
$$
$$
\cstok{\ }_{A'B'} \equiv \nabla_{F(A'} 
\; \nabla_{B')}^{\; \; \; \; F} \; ,
\eqno (3.2)
$$
we need the standard identity$^{15-17}$ 
$
\Omega_{[AB]} = {1\over 2} \epsilon_{AB} \; \Omega_{C}^{\; \; C}
$
and the spinor Ricci identities$^{8}$
$$
\cstok{\ }_{AB} \; \nu_{C}=\psi_{ABCD} \; \nu^{D}
-2 \Lambda \; \nu_{(A} \; \epsilon_{B)C} \; ,
\eqno (3.3)
$$
$$
\cstok{\ }_{A'B'} \lambda_{C'}=
{\widetilde \psi}_{A'B'C'D'} \; \lambda^{D'}
-2 \Lambda \; \lambda_{(A'} \;
\epsilon_{B')C'} \; ,
\eqno (3.4)
$$
$$
\cstok{\ }^{AB} \; \lambda_{B'}=\Phi_{\; \; \; \; M'B'}^{AB}
\; \lambda^{M'} \; ,
\eqno (3.5)
$$
$$
\cstok{\ }^{A'B'} \; \nu_{B}
={\widetilde \Phi}_{\; \; \; \; \; \; MB}^{A'B'}
\; \nu^{M} \; .
\eqno (3.6)
$$
Of course, ${\widetilde \psi}_{A'B'C'D'}$ and $\psi_{ABCD}$ are
the self-dual and anti-self-dual Weyl spinors respectively.

Thus, on using the Eqs. (2.3)-(2.8) and (3.1)-(3.6),
the basic rules of two-spinor calculus$^{15-17}$ lead to the 
compatibility equations 
$$
3 \Lambda \; \lambda_{A'}=0 \; ,
\eqno (3.7)
$$
$$
\Phi_{\; \; \; \; M'}^{AB \; \; \; C'} \; \lambda^{M'}=0 \; ,
\eqno (3.8)
$$
$$
3\Lambda \; \nu_{A}=0 \; ,
\eqno (3.9)
$$
$$
{\widetilde \Phi}_{\; \; \; \; \; \; M}^{A'B' \; \; C}
\; \nu^{M}=0 \; .
\eqno (3.10)
$$
Non-trivial solutions of (3.7)-(3.10) only exist if 
the scalar curvature and the trace-free part of the Ricci
tensor vanish. Hence the gauge transformations (2.7)-(2.8)
lead to spinor fields $\nu_{A}$ and $\lambda_{A'}$ which are
freely specifiable {\it inside} Ricci-flat backgrounds,
while the boundary conditions (1.2) are preserved under the
action of (2.7)-(2.8) providing the following conditions 
hold at the boundary:
$$
\sqrt{2} \; {_{e}n_{A}^{\; \; A'}} \;
\Bigr(\nabla^{AC'} \; \nu^{B}\Bigr)e_{BC'i}
=\pm \Bigr(\nabla^{CA'} 
\lambda^{B'}\Bigr) e_{CB'i} 
\; \; \; \; {\rm at} \; \; \; \; \partial M \; .
\eqno (3.11)
$$
\vskip 1cm
\leftline {\bf 4. Secondary Potentials in Ricci-Flat Backgrounds}
\vskip 1cm
\noindent
As shown by Penrose in Ref. 18, in a Ricci-flat manifold the 
Rarita-Schwinger potentials may be supplemented by secondary
potentials. Here we use such a construction in its local form.
For this purpose,
we introduce secondary potentials for the $\gamma$-potentials by
requiring that locally (see Ref. 18)
$$
\gamma_{A'B'}^{\; \; \; \; \; \; \; C} \equiv  \nabla_{BB'} \; 
\rho_{A'}^{\; \; \; CB} \; .
\eqno (4.1)
$$
Of course, special attention should be payed to the index ordering
in (4.1), since the primary and secondary potentials are not
symmetric (cf. Ref. 8). On inserting (4.1) into (2.3), a repeated use
of symmetrizations and anti-symmetrizations leads to the equation
(hereafter $\cstok{\ } \equiv \nabla_{CF'} \nabla^{CF'}$)
$$ \eqalignno{
\; & \epsilon_{FL} \; \nabla_{AA'} \;
\nabla^{B'(F} \; \rho_{B'}^{\; \; \; A)L}
+{1\over 2} \nabla_{\; \; A'}^{A} \;
\nabla^{B'M} \; \rho_{B'(AM)} \cr
&+ \cstok{\ }_{AM} \; \rho_{A'}^{\; \; \; (AM)}
+{3\over 8} \cstok{\ } \rho_{A'}
= 0 \; ,
&(4.2)\cr}
$$
where, following Ref. 18, we have defined
$$
\rho_{A'} \equiv \rho_{A' C}^{\; \; \; \; \; C} \; ,
\eqno (4.3)
$$
and we bear in mind that our background has to be Ricci-flat.
Thus, if the following equation holds (cf. Ref. 18):
$$
\nabla^{B'(F} \; \rho_{B'}^{\; \; \; A)L}=0 \; ,
\eqno (4.4)
$$
one finds
$$
\nabla^{B'M} \; \rho_{B'(AM)}={3\over 2} \; 
\nabla_{A}^{\; \; F'} \; \rho_{F'} \; ,
\eqno (4.5)
$$
and hence Eq. (4.2) may be cast in the form
$$
\cstok{\ }_{AM} \; \rho_{A'}^{\; \; \; (AM)}=0 \; .
\eqno (4.6)
$$
A very useful identity resulting from Eq. (4.9.13)
of Ref. 19 enables one to show that
$$
\cstok{\ }_{AM} \; \rho_{A'}^{\; \; \; (AM)}
=-\Phi_{AMA'}^{\; \; \; \; \; \; \; \; \; \; L'} \; 
\rho_{L'}^{\; \; \; (AM)} \; .
\eqno (4.7)
$$
Hence Eq. (4.6) reduces to an identity by virtue of
Ricci-flatness. Moreover, we have to insert (4.1) into the
field equation (2.4) for $\gamma$-potentials. By virtue of
(4.4) and of the identities (cf. Ref. 19)
$$
\cstok{\ }^{BM} \; \rho_{B' \; \; M}^{\; \; \; A}
=-\psi^{ABLM} \; \rho_{(LM)B'}
-\Phi_{\; \; \; \; \; B'}^{BM \; \; \; D'}
\; \rho_{\; \; MD'}^{A}
+4\Lambda \; \rho_{\; \; \; \; \; \; \; B'}^{(AB)} \; ,
\eqno (4.8)
$$
$$
\cstok{\ }^{B'F'} \; \rho_{B'}^{\; \; \; (AB)}
=3 \Lambda \; \rho^{(AB)F'}
+{\widetilde \Phi}_{\; \; \; \; \; \; \; L}^{B'F' \; \; \; A}
\; \rho_{\; \; \; \; \; \; \; B'}^{(LB)}
+{\widetilde \Phi}_{\; \; \; \; \; \; \; \; \; \; L}^{B'F'B}
\; \rho_{\; \; \; \; \; \; \; B'}^{(AL)} \; ,
\eqno (4.9)
$$
this leads to the equation
$$ 
\psi^{ABLM} \; \rho_{(LM)C'}=0 \; ,
\eqno (4.10)
$$
where we have used again the Ricci-flatness condition.

Of course, secondary potentials supplementing $\Gamma$-potentials
may also be constructed locally. On defining
$$
\Gamma_{AB}^{\; \; \; \; \; C'} \equiv
\nabla_{B'B} \; \theta_{A}^{\; \; C'B'} \; ,
\eqno (4.11)
$$
$$
\theta_{A} \equiv \theta_{AC'}^{\; \; \; \; \; C'} \; ,
\eqno (4.12)
$$
and requiring that$^{18}$
$$
\nabla^{B(F'} \; \theta_{B}^{\; \; A')L'}=0 \; ,
\eqno (4.13)
$$
one finds
$$
\nabla^{BM'} \; \theta_{B(A'M')}={3\over 2}
\nabla_{A'}^{\; \; \; F} \; \theta_{F} \; ,
\eqno (4.14)
$$
and a similar calculation yields an identity and the equation
$$ 
{\widetilde \psi}^{A'B'L'M'}
\; \theta_{(L'M')C}=0 \; .
\eqno (4.15)
$$
Note that Eqs. (4.10) and (4.15) relate explicitly the
secondary potentials to 
the curvature of the background. This inconsistency
is avoided if one of the following conditions holds:
\vskip 0.3cm
\noindent
(i) The whole conformal curvature of the background vanishes.
\vskip 0.3cm
\noindent
(ii) $\psi^{ABLM}$ and $\theta_{(L'M')C}$, or
${\widetilde \psi}^{A'B'L'M'}$ and 
$\rho_{(LM)C'}$, vanish.
\vskip 0.3cm
\noindent
(iii) The symmetric parts of the secondary potentials vanish.
\vskip 0.3cm
\noindent
In the first case one finds that the only admissible 
background is again flat Euclidean four-space with boundary,
as in Ref. 8. By contrast, in the other cases, 
left-flat, right-flat or Ricci-flat backgrounds are still
admissible, providing the secondary potentials take the form
$$
\rho_{A'}^{\; \; \; CB}=\epsilon^{CB} \; 
{\widetilde \alpha}_{A'} \; ,
\eqno (4.16)
$$
$$
\theta_{A}^{\; \; C'B'}=\epsilon^{C'B'} \;
\alpha_{A} \; ,
\eqno (4.17)
$$
where $\alpha_{A}$ and ${\widetilde \alpha}_{A'}$ solve
the Weyl equations
$$
\nabla^{AA'} \; \alpha_{A}=0 \; ,
\eqno (4.18)
$$
$$
\nabla^{AA'} \; {\widetilde \alpha}_{A'}=0 \; .
\eqno (4.19)
$$
Eqs. (4.16)-(4.19) ensure also the validity of Eqs.
(4.4), (4.13), and (A.6)-(A.7) of the appendix.

However, if one requires the preservation of Eqs. (4.4)
and (4.13) under the following gauge transformations for 
secondary potentials (the order of the indices $AL$, $A'L'$
is of crucial importance):
$$
{\widehat \rho}_{B'}^{\; \; \; AL} \equiv
\rho_{B'}^{\; \; \; AL}+\nabla_{B'}^{\; \; \; A}
\; \mu^{L} \; ,
\eqno (4.20)
$$
$$
{\widehat \theta}_{B}^{\; \; A'L'} \equiv
\theta_{B}^{\; \; A'L'}+\nabla_{B}^{\; \; A'}
\; \sigma^{L'} \; ,
\eqno (4.21)
$$
one finds compatibility conditions in Ricci-flat backgrounds
of the form
$$
\psi_{AFLD} \; \mu^{D}=0 \; ,
\eqno (4.22)
$$
$$
{\widetilde \psi}_{A'F'L'D'} \; \sigma^{D'}=0 \; .
\eqno (4.23)
$$
Thus, to ensure {\it unrestricted} gauge freedom (except at
the boundary) for the secondary potentials, one is forced
to work with flat Euclidean backgrounds. The boundary
conditions (1.2) play a role in this respect, since they make
it necessary to consider both $\psi_{i}^{A}$ and
${\widetilde \psi}_{i}^{A'}$, and hence both
$\rho_{B'}^{\; \; \; AL}$ and $\theta_{B}^{\; \; A'L'}$.
Otherwise, one might use (4.22) to set to zero the 
anti-self-dual Weyl spinor only, {\it or} (4.23) to set to
zero the self-dual Weyl spinor only, so that self-dual
(left-flat) or anti-self-dual (right-flat) Riemannian 
backgrounds with boundary would survive.
\vskip 1cm
\leftline {\bf 5. Other Gauge Transformations}
\vskip 1cm
\noindent
In the massless case, flat Euclidean backgrounds with
boundary are really the only possible choice for
spin-${3\over 2}$ potentials with a gauge freedom. 
To prove this,
we have also investigated an alternative set of gauge
transformations for primary potentials, written in
the form (cf. (2.7)-(2.8))
$$
{\widehat \gamma}_{\; \; B'C'}^{A} \equiv
\gamma_{\; \; B'C'}^{A}+\nabla_{\; \; C'}^{A} 
\; \lambda_{B'} \; ,
\eqno (5.1)
$$
$$
{\widehat \Gamma}_{\; \; \; BC}^{A'} \equiv
\Gamma_{\; \; \; BC}^{A'}+\nabla_{\; \; \; C}^{A'}
\; \nu_{B} \; .
\eqno (5.2)
$$
These gauge transformations {\it do not} correspond to the
usual formulation of the Rarita-Schwinger system, but we
will see that they can be interpreted in terms of
familiar physical concepts.

On imposing that the field equations (2.3)-(2.6) should be
preserved under the action of (5.1)-(5.2), and setting to
zero the trace-free part of the Ricci spinor (since it is
inconsistent to have gauge fields $\lambda_{B'}$ and
$\nu_{B}$ which depend explicitly 
on the curvature of the background) one finds compatibility
conditions in the form of differential equations, i.e. 
(cf. Ref. 20)
$$
\cstok{\ }\lambda_{B'}=-2\Lambda \; \lambda_{B'} \; ,
\eqno (5.3)
$$
$$
\nabla^{(A(B'} \; \nabla^{C')B)} \lambda_{B'}=0 \; ,
\eqno (5.4)
$$
$$
\cstok{\ }\nu_{B}=-2\Lambda \; \nu_{B} \; ,
\eqno (5.5)
$$
$$
\nabla^{(A'(B} \; \nabla^{C)B')} \; \nu_{B}=0 \; .
\eqno (5.6)
$$
In a flat Riemannian four-manifold with flat connection $D$,
covariant derivatives commute and $\Lambda=0$. Hence it is
possible to express $\lambda_{B'}$ and $\nu_{B}$ as
solutions of the Weyl equations
$$
D^{AB'} \; \lambda_{B'}=0 \; ,
\eqno (5.7)
$$
$$
D^{BA'} \; \nu_{B}=0 \; ,
\eqno (5.8)
$$
which agree with the flat-space version of (5.3)-(5.6).
The boundary conditions (1.2) are then preserved under the
action of (5.1)-(5.2) if $\nu_{B}$ and $\lambda_{B'}$
obey the boundary conditions (cf. (3.11))
$$
\sqrt{2} \; {_{e}n_{A}^{\; \; A'}}
\Bigr(D^{BC'} \; \nu^{A}\Bigr)e_{BC'i}
=\pm \Bigr(D^{CB'} \; \lambda^{A'}\Bigr)
e_{CB'i} \; \; \; \; {\rm at} \; \; \; \; 
{\partial M} \; .
\eqno (5.9)
$$

In the curved case, on defining
$$
\phi^{A} \equiv \nabla^{AA'} \; \lambda_{A'} \; ,
\eqno (5.10)
$$
$$
{\widetilde \phi}^{A'} \equiv \nabla^{AA'} \; \nu_{A} \; ,
\eqno (5.11)
$$
Eqs. (5.4) and (5.6) imply that 
these spinor fields solve the equations (cf. Ref. 20)
$$
\nabla_{C'}^{\; \; \; (A} \; \phi^{B)}=0 \; ,
\eqno (5.12)
$$
$$
\nabla_{C}^{\; \; (A'} \; {\widetilde \phi}^{B')}=0 \; .
\eqno (5.13)
$$
Moreover, Eqs. (5.3), (5.5) and the spinor Ricci identities
imply that
$$
\nabla_{AB'} \; \phi^{A}=2\Lambda \; \lambda_{B'} \; ,
\eqno (5.14)
$$
$$
\nabla_{BA'} \; {\widetilde \phi}^{A'}=2\Lambda \; \nu_{B} \; .
\eqno (5.15)
$$
Remarkably, the Eqs. (5.12)-(5.13) are the twistor
equations$^{15,20}$ in Riemannian four-geometries. 
The consistency conditions for the existence of non-trivial
solutions of such equations in curved four-manifolds 
are given by$^{15}$
$$
\psi_{ABCD}=0 \; ,
\eqno (5.16)
$$
and
$$
{\widetilde \psi}_{A'B'C'D'}=0 \; ,
\eqno (5.17)
$$
respectively, unless one regards $\phi^{B}$ as a four-fold
principal spinor$^{15}$ of $\psi_{ABCD}$, and
${\widetilde \phi}^{B'}$ as a four-fold principal spinor
of ${\widetilde \psi}_{A'B'C'D'}$.

Further consistency conditions for our problem are derived
by acting with covariant differentiation on the twistor
equation, i.e.
$$
\nabla_{A'}^{\; \; \; C} \; \nabla^{AA'} \; \phi^{B}
+\nabla_{A'}^{\; \; \; C} \; \nabla^{BA'} \; \phi^{A}=0 \; .
\eqno (5.18)
$$
While the complete symmetrization in $ABC$ yields (5.16),
the use of (5.18), jointly with the spinor Ricci identities
of Sec. 3, yields
$$
\cstok{\ }\phi^{B}=2\Lambda \; \phi^{B} \; ,
\eqno (5.19)
$$
and an analogous equation is found for ${\widetilde \phi}^{B'}$.
Thus, since Eq. (5.12) implies 
$$
\nabla_{C'}^{\; \; \; A} \; \phi^{B}=\epsilon^{AB}
\; \pi_{C'} \; ,
\eqno (5.20)
$$
we may obtain from (5.20) the equation 
$$
\nabla^{BA'} \; \pi_{A'}=2\Lambda \; \phi^{B} \; ,
\eqno (5.21)
$$
by virtue of the spinor Ricci identities and of (5.19).
On the other hand, in the light of (5.20), Eq. (5.14)
leads to
$$
\nabla_{AB'} \; \phi^{A}=2\pi_{B'}
=2\Lambda \; \lambda_{B'} \; .
\eqno (5.22)
$$
Hence $\pi_{A'}=\Lambda \; \lambda_{A'}$, and the 
definition (5.10) yields
$$
\nabla^{BA'} \; \pi_{A'}=\Lambda \; \phi^{B} \; .
\eqno (5.23)
$$
By comparison of (5.21) and (5.23), one gets the
equation $\Lambda \; \phi^{B}=0$. If $\Lambda \not = 0$,
this implies that $\phi^{B}$, $\pi_{B'}$ and
$\lambda_{B'}$ have to vanish, and there is no gauge
freedom fou our model. This inconsistency is avoided
if and only if $\Lambda=0$, and the corresponding 
background is forced to be totally flat, since we have
already set to zero the trace-free part of the Ricci 
spinor and the whole conformal curvature. The same argument
applies to ${\widetilde \phi}^{B'}$ and the gauge field
$\nu_{B}$. The present analysis corrects the statements
made in Sec. 8.8 of Ref. 20, where it was not realized 
that, in our massless model, a non-vanishing cosmological
constant is incompatible with a gauge freedom for the
spin-${3\over 2}$ potential. More precisely, if one sets
$\Lambda=0$ from the beginning in (5.3) and (5.5), the
system (5.3)-(5.6) admits solutions of the Weyl equation in
Ricci-flat manifolds. These backgrounds are further
restricted to be totally flat on considering the Eqs.
(4.10) and (4.15) for an arbitrary form of the secondary
potentials. As already pointed out at the end of Sec. 4,
the boundary conditions (1.2) play a role, since otherwise
one might focus on right-flat or left-flat Riemannian
backgrounds with boundary.

Yet other gauge transformations can be studied (e.g. the
ones involving gauge fields $\lambda_{B'}$ and $\nu_{B}$
which solve the twistor equations), but they are all
incompatible with a non-vanishing cosmological 
constant in the massless case.
\vskip 1cm
\leftline {\bf 6. Concluding Remarks and Open Problems}
\vskip 1cm
\noindent
The consideration of boundary conditions is essential
to obtain a well-defined formulation of physical theories
in quantum cosmology.$^{5,21,22}$ In particular,
one-loop quantum cosmology$^{3-7}$ makes it necessary to study 
spin-${3\over 2}$ potentials 
about four-dimensional Riemannian backgrounds with
boundary. The corresponding classical analysis has been performed
in our paper in the massless case, to supersede the analysis  
appearing in Refs. 8 and 23. Our results are as follows.

First, the gauge transformations (2.7)-(2.8) are compatible with the
massless Rarita-Schwinger equations providing the 
background four-geometry is Ricci-flat. However, the presence
of a boundary restricts the gauge freedom, since the boundary
conditions (1.2) are preserved under the action of (2.7)-(2.8)
only if the boundary conditions (3.11) hold.

Second, the Penrose construction of secondary potentials in Ricci-flat
four-manifolds shows that the admissible backgrounds may be
further restricted to be totally flat, or left-flat, or 
right-flat, unless these secondary potentials take the special 
form (4.16)-(4.17).
Hence the secondary potentials 
supplementing the Rarita-Schwinger potentials have a very clear 
physical meaning in Ricci-flat four-geometries with boundary:
they are related to the spinor fields $\Bigr(\alpha_{A},
{\widetilde \alpha}_{A'}\Bigr)$ corresponding to the Majorana
field in the Lorentzian version of (2.3)-(2.6). [One should
bear in mind that, in real Riemannian 
four-manifolds, the only admissible
spinor conjugation is Euclidean conjugation, which is anti-involutory
on spinor fields with an odd number 
of indices.$^{5,20,24}$ Hence no Majorana
field can be defined in real Riemannian four-geometries]

Third, to ensure unrestricted gauge freedom for the secondary
potentials, one is forced to work with flat Euclidean
backgrounds, when the boundary conditions (1.2) are imposed.
Thus, the very restrictive results obtained in Refs. 8 and
23 for massless Dirac potentials with the boundary conditions
(1.1) are indeed confirmed also for massless Rarita-Schwinger 
potentials subject to the supersymmetric boundary conditions
(1.2). Interestingly, a formalism originally
motivated by twistor theory$^{15,18,20,23-26}$ 
has been applied to classical
boundary-value problems relevant for one-loop quantum cosmology.

Fourth, the gauge transformations (5.1)-(5.2) with non-trivial
gauge fields are compatible with the field equations 
(2.3)-(2.6) if and only if the background is totally flat.
The corresponding gauge fields solve the Weyl equations
(5.7)-(5.8), subject to the boundary conditions (5.9).
Indeed, it is well-known that the Rarita-Schwinger 
description of a massless spin-${3\over 2}$ field is
equivalent to the Dirac description in a special choice of
gauge.$^{18}$ In such a gauge, the spinor fields 
$\lambda_{B'}$ and $\nu_{B}$ solve the Weyl equations,
and this is exactly what we find in Sec. 5 on choosing
the gauge transformations (5.1)-(5.2).

A non-vanishing cosmological constant can be consistently
studied when a {\it massive} spin-${3\over 2}$ potential
is studied.$^{27}$ For this purpose, one has to replace the
spinor covariant derivative $\nabla_{AA'}$ in the field
equations (2.3)-(2.6) by a new spinor covariant derivative
$S_{AA'}$ which reduces to $\nabla_{AA'}$ when $\Lambda=0$.
In the language of $\gamma$-matrices, one has
(cf. Ref. 27)
$$
S_{\mu} \equiv \nabla_{\mu}+f(\Lambda)\gamma_{\mu} \; ,
\eqno (6.1)
$$
where $f(\Lambda)$ vanishes at $\Lambda=0$, and $\gamma_{\mu}$
are the $\gamma$-matrices. We are currently investigating the
reformulation of Secs. 2-5 in terms of the definition (6.1).
In particular, it appears interesting to understand,
by using two-spinor formalism, whether twistors can
generate the gauge freedom for a class of {\it massive} 
spin-${3\over 2}$ potentials in conformally flat 
Einstein four-geometries with boundary.
Moreover, other interesting problems are found to arise:
\vskip 0.3cm
\noindent
(i) Can one relate Eqs. (4.4) and (4.13) to the theory of
integrability conditions relevant for massless fields in
curved backgrounds (see Ref. 18 and our appendix) ?
What happens when such equations do not hold ?
\vskip 0.3cm
\noindent
(ii) Is there an underlying global theory of Rarita-Schwinger
potentials ? In the affirmative case, what are the key features
of the global theory ? 
\vskip 0.3cm
\noindent
(iii) Can one reconstruct the Riemannian four-geometry from the
twistor space in Ricci-flat or conformally flat backgrounds with
boundary, or from whatever is going to replace twistor
space ?

Thus, the results and problems presented in our paper seem to add
evidence in favour of a deep link existing between twistor
geometry, quantum cosmology and modern field theory.
\vskip 1cm
\leftline {\bf Acknowledgments}
\vskip 1cm
\noindent
G. Esposito is much indebted to Roger Penrose for making it possible
for him to study the earliest version of his work on secondary
potentials which supplement Rarita-Schwinger potentials.
Our joint paper was supported in part by the European Union under
the Human Capital and Mobility 
Programme, and by research funds of the
Italian Ministero per l'Universit\`a e la Ricerca Scientifica
e Tecnologica. Moreover,
the research described in this publication was made possible in
part by Grant No MAE300 from the International Science Foundation
and from the Russian Government.
The work of A. Kamenshchik was partially supported by the
Russian Foundation for Fundamental Researches through grant No
94-02-03850-a, and by the Russian Research Project 
``Cosmomicrophysics". A. Kamenshchik is grateful to the Dipartimento
di Scienze Fisiche dell'Universit\`a di Napoli and to the
Istituto Nazionale di Fisica Nucleare for kind hospitality and
financial support during his visit to Naples in May 1994.
\vskip 10cm
\leftline {\bf Appendix}
\vskip 1cm
\noindent
Following Ref. 13, one can locally express the $\Gamma$-potentials 
of (2.1) as (cf. (4.11))
$$
\Gamma_{\; \; BB'}^{A} \equiv \nabla_{BB'} \; \alpha^{A} \; .
\eqno (A.1)
$$
Thus, acting with $\nabla_{CC'}$ on both sides of (A.1), 
symmetrizing over $C'B'$
and using the spinor Ricci identity (3.6), one finds
$$
\nabla_{C(C'} \; \Gamma_{\; \; \; \; \; B')}^{AC}
={\widetilde \Phi}_{B'C'L}^{\; \; \; \; \; \; \; \; \; \; A} \;
\alpha^{L} \; .
\eqno (A.2)
$$
Moreover, acting with $\nabla_{C}^{\; \; C'}$ on both sides of (A.1),
putting $B'=C'$ (with contraction over this index), and using the
spinor Ricci identity (3.3) leads to
$$
\epsilon^{AB} \; \nabla_{(C}^{\; \; \; C'} \; 
\Gamma_{\mid A \mid B)C'}=-3\Lambda \; \alpha_{C} \; .
\eqno (A.3)
$$
Eqs. (A.1)-(A.3) rely on the conventions in Ref. 13.   
However, to achieve agreement with the conventions in Ref. 18
and in our paper, the Eqs. (2.3)-(2.6) are obtained
by defining (for the effect of
torsion terms, see comments following Eq. (21) in Ref. 13)
$$
\Gamma_{B \; \; \; \; B'}^{\; \; \; A}
\equiv \nabla_{BB'} \; \alpha^{A} \; ,
\eqno (A.4)     
$$
$$
\gamma_{A' \; \; \; \; C}^{\; \; \; B'}
\equiv \nabla_{CA'} \; {\widetilde \alpha}^{B'} \; .
\eqno (A.5)
$$
On requiring that (A.5) and (4.1) should agree,
one finds by comparison that 
$$
\nabla_{BB'} \; \rho_{A'}^{\; \; \; (CB)}
=2 \nabla_{\; \; [A'}^{C} \;
{\widetilde \alpha}_{B']} \; ,
\eqno (A.6)
$$
which is obviously satisfied if $\rho_{A'}^{\; \; \; (CB)}=0$
and ${\widetilde \alpha}_{B'}$ obeys the Weyl equation (4.19).
Similarly, by comparison of (A.4) and (4.11) one finds
$$
\nabla_{B'B} \; \theta_{A}^{\; \; (C'B')}
=2 \nabla_{\; \; \; [A}^{C'} \; \alpha_{B]} \; ,
\eqno (A.7)
$$
which is satisfied if Eqs. (4.17)-(4.18) hold.

In the original approach by Penrose,$^{18}$ one describes
Rarita-Schwinger potentials in flat space-time in terms of a
rank-three vector bundle with local coordinates 
$\Bigr(\eta_{A},\zeta \Bigr)$, and an operator $\Omega_{AA'}$
whose action is defined by
$$
\Omega_{AA'}(\eta_{B},\zeta) \equiv
\Bigr({\cal D}_{AA'}\eta_{B},{\cal D}_{AA'}\zeta
-\eta^{C}\rho_{A'AC}\Bigr) \; ,
\eqno (A.8)
$$
$\cal D$ being the flat Levi-Civita connection of
Minkowski space-time. The gauge transformations are then
$$
\Bigr({\widehat \eta}_{B},{\widehat \zeta}\Bigr) 
\equiv \Bigr(\eta_{B},\zeta+\eta_{A}\xi^{A}\Bigr) \; ,
\eqno (A.9)
$$
$$
{\widehat \rho}_{A'AB} \equiv \rho_{A'AB}
+{\cal D}_{AA'}\xi_{B} \; .
\eqno (A.10)
$$
For the operator defined in (A.8), the integrability
condition on $\beta$-planes$^{20}$ turns out to be
$$
{\cal D}^{A'(A} \; \rho_{A'}^{\; \; \; B)C}=0 \; .
\eqno (A.11)
$$
It now remains to be seen whether, at least in Ricci-flat
backgrounds, an operator can be defined (cf. (A.8)) whose
integrability condition on $\beta$-surfaces$^{20}$ is indeed
given by Eq. (4.4) (cf. Eq. (A.11)).
\vskip 10cm
\leftline {\bf References}
\vskip 1cm
\item {1.}
H. C. Luckock and I. G. Moss, {\it Class. Quantum Grav.}
{\bf 6}, 1993 (1989).
\item {2.}
H. C. Luckock, {\it J. Math. Phys.} {\bf 32}, 1755 (1991).
\item {3.}
P. D. D'Eath and G. Esposito, {\it Phys. Rev.} {\bf D43},
3234 (1991).
\item {4.}
G. Esposito, {\it Int. J. Mod. Phys.} {\bf D3}, 593 (1994).
\item {5.}
G. Esposito, {\it Quantum Gravity, Quantum Cosmology and
Lorentzian Geometries}, Lecture Notes in Physics, New
Series m: Monographs, Volume m12 (Springer-Verlag, 
Berlin, 1994).
\item {6.}
A. Y. Kamenshchik and I. V. Mishakov, {\it Phys. Rev.}
{\bf D47}, 1380 (1993).
\item {7.}
A. Y. Kamenshchik and I. V. Mishakov, {\it Phys. Rev.}
{\bf D49}, 816 (1994).
\item {8.}
G. Esposito and G. Pollifrone, {\it Class. Quantum Grav.}
{\bf 11}, 897 (1994).
\item {9.}
P. Breitenlohner and D. Z. Freedman, {\it Ann. Phys.}
{\bf 144}, 249 (1982).
\item {10.}
S. W. Hawking, {\it Phys. Lett.} {\bf B126}, 175 (1983).
\item {11.}
P. D. D'Eath, {\it Phys. Rev.} {\bf D29}, 2199 (1984).
\item {12.}
A. Sen, {\it J. Math. Phys.} {\bf 22}, 1781 (1981).
\item {13.}
P. C. Aichelburg and H. K. Urbantke, {\it Gen. Rel. Grav.}
{\bf 13}, 817 (1981).
\item {14.}
R. Penrose, {\it Twistor Newsletter} {\bf 33}, 1 (1991).
\item {15.}
R. Penrose and W. Rindler, {\it Spinors and Space-Time, Vol. 2:
Spinor and Twistor Methods in Space-Time Geometry} 
(Cambridge University Press, Cambridge, 1986).
\item {16.}
R. S. Ward and R. O. Wells, {\it Twistor Geometry and Field
Theory} (Cambridge University Press, Cambridge, 1990). 
\item {17.}
J. M. Stewart, {\it Advanced General Relativity}
(Cambridge University Press, Cambridge, 1991).
\item {18.}
R. Penrose, in {\it Twistor Theory}, ed. S. Huggett
(Marcel Dekker, New York, 1994).
\item {19.}
R. Penrose and W. Rindler, {\it Spinors and Space-Time, Vol. 1:
Two-Spinor Calculus and Relativistic Fields} 
(Cambridge University Press, Cambridge, 1984).
\item {20.}
G. Esposito, {\it Complex General Relativity}, Fundamental
Theories of Physics, Volume 69 (Kluwer, Dordrecht, 1995).
\item {21.}
J. B. Hartle and S. W. Hawking, {\it Phys. Rev.}
{\bf D28}, 2960 (1983).
\item {22.}
S. W. Hawking, {\it Nucl. Phys.} {\bf B239}, 257 (1984).
\item {23.}
G. Esposito and G. Pollifrone, in
{\it Twistor Theory}, ed. S. Huggett 
(Marcel Dekker, New York, 1994).
\item {24.}
N. M. J. Woodhouse, {\it Class. Quantum Grav.}
{\bf 2}, 257 (1985).
\item {25.}
L. J. Mason and R. Penrose, {\it Twistor Newsletter} 
{\bf 37}, 1 (1994).
\item {26.}
J. Frauendiener, {\it Twistor Newsletter} 
{\bf 37}, 7 (1994).
\item {27.}
P. K. Townsend, {\it Phys. Rev.} {\bf D15}, 2802 (1977).

\bye